# A Theoretical and Empirical Evaluation of Software Component Search Engines, Semantic Search Engines and Google Search Engine in the Context of COTS-Based Development

Nacim YANES, Sihem BEN SASSI, and Henda HAJJAMI BEN GHEZALA

*Abstract*—COTS-based development is a component reuse approach promising to reduce costs and risks, and ensure higher quality. The growing availability of COTS components on the Web has concretized the possibility of achieving these objectives. In this multitude, a recurrent problem is the identification of the COTS components that best satisfy the user requirements. Finding an adequate COTS component implies searching among heterogeneous descriptions of the components within a broad search space. Thus, the use of search engines is required to make more efficient the COTS components identification. In this paper, we investigate, theoretically and empirically, the COTS component search performance of eight software component search engines, nine semantic search engines and a conventional search engine (Google). Our empirical evaluation is conducted with respect to precision and normalized recall. We defined ten queries for the assessed search engines. These queries were carefully selected to evaluate the capability of each search engine for handling COTS component identification.

*Index Terms*— COTS identification, keyword-based search engine, software component search engine, semantic search engine, semantic search performance, evaluation

## I. INTRODUCTION

COTS-based development (CBD) is changing the way information systems are being developed and delivered to the end user. It consists in building applications by selecting and integrating pre-packaged solutions, known as Commercial-Off-The-Shelf (COTS) components. It is common that researchers and practitioners use the word "COTS" with different meanings. Some of them use the term COTS covering freeware and Open Source Software as well as other kinds of components; whilst others are more restrictive [1]. In order to be precise, in this paper we follow the Software Engineering Institute definition: "A COTS product is a [software] product that is: (1) sold, leased, or licensed to the general public; (2) offered by a vendor trying to profit from it; (3) supported and evolved by the vendor, who retains the intellectual property rights; (4) available in multiple, identical copies; and (5) used without source code modification by a consumer" [2]..

Several advantages of COTS-based development have been identified for both component providers (COTS vendors) and component consumrs (COTS integrators). Component providers have the opportunity to enter new markets thanks to a huge marketplace already available and steadily growing, both in the variety of segments and the COTS components offered therein. In fact, numerous on-line companies such as componentsource.com and componentace.com are now selling software components and related services [3]. These companies not only allow end users to buy their own components, but also act as intermediaries, providing access to third-party products. On the other hand, components consumers can increase development productivity, shorten development life cycle, increase quality of the final system, and decrease development costs. They have also the flexibility to quickly substitute COTS components with newer ones, containing additional features, in order to respond to competitive forces and changing market conditions [4].

A typical process for COTS-based development consists in five main steps including searching and identifying COTS components candidates, selecting the most appropriate ones and assembling them with the other components [5]. Identifying COTS components is a critical activity having to cope with some challenging marketplace characteristics related to its widespread, evolvable and growing nature. Finding adequate COTS components involves searching among several on-line commercial repositories maintained by COTS vendors and publishers. Thus, the use of search engines is required to enhance the COTS identification effectiveness and efficiency.

Publicly available search engines on the Web are generally keyword-based search engines. These search engines, such as Google and Yahoo, have a large user-base. Besides, some specialized search engines have been proposed to improve search results. These particular search engines search just for information in a particular topic or category on the Web [6]. On the other hand, the research trend on semantic search engines is increasing. In fact, researchers have conducted several studies and have proposed search engines, methods and technologies for semantic search. Therefore, evaluation of search performance of keyword-based search engines, software components search engines and semantic search engines is a valuable work intended to assess to which extent they are able to satisfy COTS component users.

The remainder of this paper is structured as follows: section 2 presents our study sample of search engines assessed in the context of COTS-based development. Section 3 focuses on the theoretical evaluation of the identified search engines in order to assess their success in the identification of COTS components marketed on the Web. Section 4 describes first the methodology employed to empirically evaluate the search engines in terms of precision and normalized recall. Then, it reports and discusses the experimental findings. Finally, the paper ends with a conclusion and work in progress.

## II. THE STUDY SAMPLE OF SEARCH ENGINES

Publicly available search engines on the Web are generally keyword-based search engines. These search engines, such as Google, have a large user-base. Besides, some specialized search engines are available on the web allowing search and retrieval of software components. We studied a sample of these software component search engines in order to assess their success in the identification of COTS components marketed on the Web.

On the other hand, the research trend on semantic search engines is increasing. In fact, researchers have conducted several studies and have proposed search engines, methods and technologies for semantic search. Therefore, evaluation of search performance of semantic search engines is a valuable work intended to assess to which extent they are able to satisfy COTS component users.

We used the Alexa[1] Web Company to identify a set of software component search engines and semantic search engines to which we added other ones to form our sample. This latter comprises, as shown in Table I and Table II, eight software component search engines, nine semantic search engine and a traditional search engine namely Google.

TABLE I
ASSESSED SOFTWARE COMPONENT SEARCH ENGINES

| Software component search engine | URL |
|---|---|
| Capterra | www.capterra.com |
| Downseek | www.downseek.com |
| Software Source | http://www.peoplemanagement.co.uk/hrsuppliers/listing/guide/hr-software |
| CNET | www.cnet.com |
| Quelsoft | www.quelsoft.com |
| Libellules | www.libellules.ch |
| Logitheque | www. Logitheque.com |
| Freeshareweb | http://minelog.com/blog/ |

[1] http://www.alexa.com

TABLE II
ASSESSED SEMANTIC SEARCH ENGINES & GOOGLE

| Semantic search engine | URL |
|---|---|
| Hakia | www.hakia.com |
| Yauba | www.yauba.com |
| SenseBot | www.sensebot.net |
| Powerset | www.powerset.com |
| Cognition | www.cognition.com |
| Lexxe | www.lexxe.com |
| Exalead | www.exalead.com/search |
| FactBites | www.factbites.com |
| Kosmix | www.kosmix.com |
| Google | www.google.com |

## III. THEORETICAL EVALUATION OF SEARCH ENGINES

In this section, we present the theoretical study carried out to assess the software components search engines, semantic search engines and Google in the context of COTS-based development.

### A. Assessment Criteria

We used criteria classified on four categories. Some of the criteria are inspired from those we have defined in [6], and to which we added new ones.

*Search Criteria*
1) Search Method: It includes four values which are: search by category (c); search by keywords (kw); search by category and keywords (c&kw); and metasearching (ms).
2) Advanced Search: It describes if the search engine provides any advanced mechanism to perform the search. It can be Boolean operators or refined searches.
3) Semantic Search: It describes if the search engine provides any mechanism to describe knowledge about the COTS component and to reason on this knowledge.

*Search Results Criteria*
1) Component type: It describes the type of software components searched. The component type can be COTS component (COTS), shareware (Sh) or freeware (Fr).
2) Result limited to software component: it indicates if the result contains only software component or includes other information such as articles about software components, technical reports, case studies, etc.
3) Rank mechanism: It specifies if the search engine provides a rank mechanism allowing sorting results according to users' preferences.

*Indexation Criteria*
1) Portals it relies on: It describes the document repositories in which the search engine relies on.
2) Indexation: It refers if the indexing of components is intended to be automatic or manual.
3) Nature of searched documents: It describes the nature of indexed and searched Web documents, i.e. structured documents (STR), semi-structured and/or unstructured documents (UNSTR).

*Other Criteria*
1) Language: It describes the language(s) supported by the COTS search engine.
2) Personalization: It specifies if the search engine provides any personalization mechanism taking into consideration the users preferences and their domains of interest during search process.

Note that some assessment criteria can be applied only on software component search engines. These criteria are "semantic search", "component type" and "result limited to software component". On the other hand, the "nature of searched documents" and "portals it relies on" criteria are used only for evaluating semantic search engines and Google.

*B. Assessment Results Regarding Software Component Search Engines*

Table III, Table IV and Table V summarize the assessment results of the eight software component search engines according to the defined criteria.

In fact, the study revealed that components specialized search engines manage small software components indexes which make easier the searching task. Indeed, users have not to browse long document lists.

TABLE III
RESULTS BASED ON SEARCH CRITERIA

| Software components Search Engines | Search Method | Semantic Search | Advanced Search | |
|---|---|---|---|---|
| | | | Boolean | Refinement |
| Capterra | c&kw | – | ✓ | – |
| DownSeek | c&kw | – | ✓ | – |
| Software Source | c&kw | – | ✓ | – |
| CNET | Kw | – | ✓ | – |
| Quelsoft | c&kw | – | – | – |
| Libellules | c | – | – | – |
| Logitheque | kw | – | – | – |
| Freeshareweb | c&kw | – | – | – |

However, several shortcomings have been detected following our theoretical evaluation. These shortcomings are described in the sequel:

1) Absence of refinement options: The assessed specialized search engines, as illustrated in table III, don't take into account the specific characteristics of COTS components (target platform, vendor, etc.).
2) Absence of semantic search: The totality of the software component search engines are based on a basic syntactic search as shown in Table III. None of the assessed search engines has integrated the semantic aspect in order to improve the results relevance.
3) Component type supported: Few search engines focus only on commercial software components. In fact, Table IV illustrates that the majority of these search engines look for freewares and sharewares. In addition, some assessed search engines include other information such as articles about software components, technical reports and case studies. These information are useless in this CBD step and irrelevant to users searching for software components.
4) Language supported: To significantly improve the search experience, a search engine should support many languages. However, Table V shows that the majority of the assessed software components engines do not have this ability. As a matter of fact, they don't take into account both French and English languages.

TABLE IV
RESULTS BASED ON SEARCH RESULTS CRITERIA

| Software components Search Engines | Search results | | |
|---|---|---|---|
| | Component type | Result limited to software component | Rank mechanism |
| Capterra | COTS, Fr | – | – |
| DownSeek | Sh, Fr | ✓ | – |
| Software Source | COTS | ✓ | – |
| CNET | Sh | ✓ | – |
| Quelsoft | COTS | ✓ | – |
| Libellules | Sh, Fr | ✓ | – |
| Logitheque | COTS, Sh, Fr | ✓ | ✓ |
| Freeshareweb | Sh, Fr | ✓ | ✓ |

5) Manual indexation: The majority of the assessed specialized search engines use a manual indexing of software components which makes the contents of their indexes limited. Besides, they don't provide a comprehensive description of indexed software components. In fact, indexed information about components are limited to their name, textual description of their functionalities and their supplier name. They don't include specific characteristics of software components which are important to accomplish the others steps of CBD such as evaluation and integration steps.

TABLE V
RESULTS BASED ON INDEXATION AND OTHER CRITERIA

| Software components Search Engines | Indexation | Language | Personalization |
|---|---|---|---|
| Capterra | Manual | English | – |
| DownSeek | Manual | English | – |
| Software Source | Manual | English | – |
| CNET | Manual | English | – |
| Quelsoft | Manual | French | – |
| Libellules | Manual | French | – |
| Logitheque | Manual | French | – |
| Freeshareweb | Manual | English, French | – |

6) Lack of personalization mechanism: Table V illustrates that none of the assessed software component search engines focus on representation and exploitation of users' preferences and intentions during search process. In fact, when using these search engines, people who input the same keywords at the same time will get exactly the same list of results.

*C. Assessment Results Regarding Semantic Search Engines and Google*

Table VI, Table VII and Table VIII summarize the assessment results of the nine semantic search engines identified in the practice and Google.

Several limitations were detected following our study of the semantic search engines and Google:

1) Absence of refinement options: Similarly to the software component search engines, the majority of semantic search engines do not allow users to refine their queries. Indeed, our study revealed that only Cognition, Lexxe and Google provide refinement options. However, these refinement options don't take into account the specific characteristics of COTS components (target platform, vendor, etc.).

TABLE VI
RESULTS BASED ON SEARCH CRITERIA

| Semantic Search Engines | Search Method | Advanced Search | |
|---|---|---|---|
| | | Boolean | Refinement |
| Hakia | Kw | – | – |
| Yauba | Kw | ✓ | – |
| SenseBot | Ms | – | – |
| Powerset | Kw | – | – |
| Cognition | Ms | ✓ | ✓ |
| Lexxe | Kw | ✓ | ✓ |
| Exalead | Kw | ✓ | – |
| FactBites | kw | – | – |
| Kosmix | C&Kw | ✓ | – |
| Google | C&Kw | ✓ | ✓ |

2) Nature of searched documents: Table VII shows that some semantic search engines index only structured documents such as Wikipedia articles. However, the largest volume of Web documents is unstructured. Therefore, the retrieving process of these search engines could return zero results in response to a submitted query, although relevant documents to this latter are available on the Web.

3) Lack of personalized mechanism: None of the assessed semantic search engines, as shown in Table VIII, focus on representation and exploitation of users' preferences and intentions during search process. For example, when we submitted the query "Commercial text processing software", the majority of semantic search engines return in their results software components about text mining tools. These results could be relevant if we were looking for text mining tools; however, our intention was to get text editors in result. Whatever the user's domains of interest, all the assessed semantic search engines return the same list of results. We encounter the same problem when using Google. In fact, people around the world who input the same keywords at the same time will get exactly the same search result regardless of their past search history on Google.

4) Language supported: The diversity of the internet is reflected not only in its users, information formats and information content, but also in the languages used. As more and more information becomes available in different languages, multiple language support in a search engine becomes more important. Nevertheless, Table VIII shows that only few semantic search engines support several languages including English and French.

TABLE VII
RESULTS BASED ON INDEXATION CRITERIA

| Semantic Search Engines | Indexation | Nature of searched documents | Portals it relies on |
|---|---|---|---|
| Hakia | Automatic | STR & UNSTR | Entire Web |
| Yauba | Automatic | STR & UNSTR | Entire Web |
| SenseBot | Automatic | STR & UNSTR | Google, Yahoo and Bing |
| Powerset | Automatic | STR | Wikipedia and Freebase |
| Cognition | Automatic | STR | Public.resource.org, MEDLINE, Gospels and Wikipedia. |
| Lexxe | Automatic | STR & UNSTR | Entire Web |
| Exalead | Automatic | STR & UNSTR | Entire Web |
| FactBites | Automatic | STR & UNSTR | Entire Web |
| Kosmix | Automatic | STR & UNSTR | Entire Web |
| Google | Automatic | STR & UNSTR | Entire Web |

TABLE VIII
RESULTS BASED ON SEARCH RESULTS AND OTHER CRITERIA

| Semantic Search Engines | Rank Mechanism | Personalization | Language |
|---|---|---|---|
| Hakia | – | – | English |
| Yauba | – | – | Many languages among other English and French |
| SenseBot | – | – | English, French, German and Spanish |
| Powerset | ✓ | – | English |
| Cognition | ✓ | – | English |
| Lexxe | – | – | English |
| Exalead | – | – | English, French and Spanish |
| FactBites | – | – | English |
| Kosmix | – | – | English |
| Google | – | – | Many languages among other English and French |

## IV. EMPIRICAL EVALUATION OF SEARCH ENGINES

This section describes first the methodology employed to evaluate search engines in terms of precision and normalized recall. Then, it reports and discusses the experimental findings.

### A. Method

Initially, we carefully chose ten queries that describe COTS components in various application domains as shown in Table IX. Then, these ten queries were run on each of the selected search engines.

TABLE IX
QUERY LIST

| Query number | Query |
|---|---|
| Q1 | Commercial Firewall software |
| Q2 | Commercial email address verification software |
| Q3 | Commercial credit card authorization software |
| Q4 | Commercial image compression software |
| Q5 | Commercial Web site development |
| Q6 | Commercial HTML editing software |
| Q7 | Commercial text processing software |
| Q8 | Commercial order management software |
| Q9 | Commercial Enterprise Resource Planning software |
| Q10 | Commercial Human resource planning software |

After each run of the query, the first fifty documents retrieved were evaluated using binary human relevance judgment. Total 6559 documents were evaluated by the same author. Moreover, all the searches and evaluations were performed in minimal non-distant time slots in order to have stable performance measurement of search engines.

*Evaluation of Performance*

Our relevance judgment was binary; which means that every document was classified as "relevant" or "non relevant". We respected the following criteria in order to evaluate the retrieved documents: 1) Documents that contain either a description or a link to COTS components satisfying the searched query were considered "relevant"; 2) documents that contain a description of free and open source component satisfying the searched query were considered "non relevant"; 3) documents having same content but originating from different Web addresses were considered as different ones [7]; 4) in case of document duplication, only the first document that was retrieved was assessed; and 5) if the retrieved document was not accessible, then it was classified as "non relevant" one.

*Performance Measurement*

The effectiveness measurements in information retrieval are typically of precision and recall. Precision is defined as the ratio of the number of relevant documents retrieved to the number of total documents retrieved [8]. Recall is defined as the ratio of the number of relevant documents retrieved to the number of total relevant documents. It has been a common practice in the evaluation of search engines to exclude recall for obvious reason [7], though there were some recommendations in the literature for estimating normalized recall [9].

Precision and normalized recall ratios of the assessed search engines were calculated at various cut-off points (first 10, 20, 30, 40 and 50 documents retrieved) for each pair of query and search engine. The precision at different cut-offs can be used to roughly see how scores of relevant documents are distributed over their ranks [7]. Note that, in our evaluation, when the number of documents retrieved is smaller than the cut-off point at the hand, precision was calculated over total documents retrieved. To calculate normalized recall at various cut-off points, we used the formula proposed by Bollmann et al. [9]. The Rnorm is defined as:

$$R_{norm} = 1 - (\Sigma^n_{i=1} r_i - \Sigma^n_{i=1} i) / n*(N\_n) \quad (1)$$

Where n is the number of relevant document in the distribution, N is the number of documents in the distribution, ri is the rank of the ith relevant document.

### B. Experiment Results

Findings of our empirical evaluation and the results analysis are discussed below.

*The number of Documents Retrieved by Search Engines*

We used the number of zero retrievals (i.e., no documents retrieved) or retrievals containing no relevant documents (i.e., the precision ratio is zero) to assess the retrieval performance of search engines [8]. The number of relevant documents retrieved for each query by each search engine for the first fifty documents retrieved is shown in Table X and Table XI. The row labeled "Total Relevant" shows the total number of relevant documents retrieved; the second row labeled "Total Retrieved" shows the total number of documents retrieved by each assessed search engine and the third one labeled "Average" shows the mean number per search query.

According to our experiment results concerning the semantic search engines and Google, Cognition and Factbites could not retrieve any relevant document for 5 of 10 queries. Likewise, Powerset could not retrieve any relevant document for 2 of 10 queries. The same figure was 1 out of 10 for Sensebot and Kosmix. On the other hand, if we examine both zero retrievals and retrievals with no relevant documents, Cognition and Factbites could not retrieve any relevant document for 6 of 10 queries. However, Hakia, Yauba, Lexxe, Exalead, and Google retrieved at least one relevant document for each query.

As illustrated by Table X, Lexxe and Factbites retrieved the highest and the lowest numbers of relevant documents for 10 queries, respectively. Total number of documents retrieved by semantic search engines was 3557, of which 307 were relevant. The average precisions of semantic search engines are between 4% and 20%. In other words, the best semantic search engine among the assessed ones returns 8 non relevant in 10 retrieved documents. On the other hand, Google mean number of relevant documents per query was higher than those of all assessed semantic search engines.

TABLE X
NUMBER OF RELEVANT DOCUMENTS RETRIEVED BY SEMANTIC SEARCH ENGINES AND GOOGLE

| Query Number | Hakia | Yauba | Sensebot | Powerset | Cognition | Lexxe | Exalead | Factbites | Kosmix | Google |
|---|---|---|---|---|---|---|---|---|---|---|
| Q1 | 13/50 | 14/50 | 4/50 | 9/50 | 8/50 | 9/50 | 7/50 | 3/28 | 7/18 | 16/50 |
| Q2 | 6 /50 | 4/50 | 2/13 | 1/50 | 0/3 | 3/19 | 2/50 | 0/0 | 5/18 | 18/50 |
| Q3 | 4/50 | 4/50 | 1/50 | 0/50 | 0/0 | 4/50 | 2/50 | ½ | 3/17 | 16/50 |
| Q4 | 3/50 | 5/50 | 1/50 | 1/50 | 0/50 | 7/50 | 4/50 | 0/13 | 0/18 | 19/50 |
| Q5 | 6/50 | 5/50 | 1/50 | 1/50 | 2/50 | 3/50 | 2/50 | 2/31 | 1/18 | 14/50 |
| Q6 | 7/50 | 8/50 | 1/11 | 2/50 | 0/4 | 9/50 | 2/50 | 0/31 | 2/18 | 16/50 |
| Q7 | 2/50 | 2/50 | 0/17 | 2/50 | 0/50 | 3/50 | 2/50 | 0/22 | 1/18 | 17/50 |
| Q8 | 3/50 | 5/50 | 2/50 | 1/50 | 0/50 | 7/50 | 3/50 | 0/31 | 7/18 | 23/50 |
| Q9 | 4/50 | 7/50 | 3/50 | 5/50 | 7/50 | 4/50 | 7/50 | 1/7 | 6/17 | 22/50 |
| Q10 | 4/50 | 6/50 | 1/50 | 0/50 | 1/46 | 15/50 | 1/50 | 0/2 | 4/17 | 21/50 |
| Total Relevant | 52 | 60 | 16 | 22 | 18 | 64 | 32 | 7 | 36 | 182 |
| Total Retrieved | 500 | 500 | 391 | 500 | 353 | 469 | 500 | 167 | 177 | 500 |
| Average | 5.2 | 6.0 | 1.6 | 2.2 | 1.8 | 6.4 | 3.2 | 0.7 | 3.6 | 18.2 |

As a consequence, we notice that even semantic search engines are meaning-based; they are less effective than Google in the context of COTS component search.

Regarding software component search engines, only Downseek retrieved at least one relevant document for each query. Freeshareweb and Libellules could not retrieve any document for 4 of 10 queries. Likewise, Software Source, Quelsoft, and Logitheque could not retrieve any document for 2 of 10 queries. On the other hand, Cnet and Libellules could not retrieve any relevant document for 3 of 10 queries. The same figure was 1 out of 10 for Software Source, Capterra and Freeshareweb.

As Table XI shows, Capterra and Libellules retrieved the highest and the lowest numbers of relevant documents for 10 queries, respectively. Total number of documents retrieved by software component search engines was 2502, of which 480 were relevant. To put it differently, about 8 in 10 documents retrieved by software component search engines were not relevant. Consequently, although these search engines are specific to reusable software components, they break down in the context of COTS component identification.

*Precision Ratios*

Mean precision values of the assessed search engines in various cut-off points (for first 10, 20, 30, 40, and 50 documents retrieved) are shown in Fig. 1 and Fig. 2.

Our first observation is that all the assessed search engines have low precision ratios. In fact, although submitted queries are about commercial software components, the assessed search engines display free software components and unrelated documents in their results.

Regarding the semantic search engines, Hakia has the highest precision ratios at cut-off point 10 (28%). With increase in cut-off point value, the precision ratio of Hakia, Yauba, Sensebot, Powerset, Lexxe, and Factbites decreased for all cut-off points.

Generally, precision ratios of search engines decreased when the cut-off point values were increased. Exalead's precision ratio decreased at cut-off points 10, 20, 30, and 40 then increased slightly at cut-off point 50. Kosmix has zero as precision ratio at cut-off points 30, 40, and 50 since the maximum number of retrieved documents for each search query was 18. Similarly, Factbites has zero as precision ratio at cut-off point 50 as it returned a maximum of 31 documents per query.

Based on Fig. 1, we can classify the assessed semantic search engines into two classes according to their precision ratios. The first class includes the semantic search engines with the highest ratios, namely Hakia, Lexxe, and Yauba.

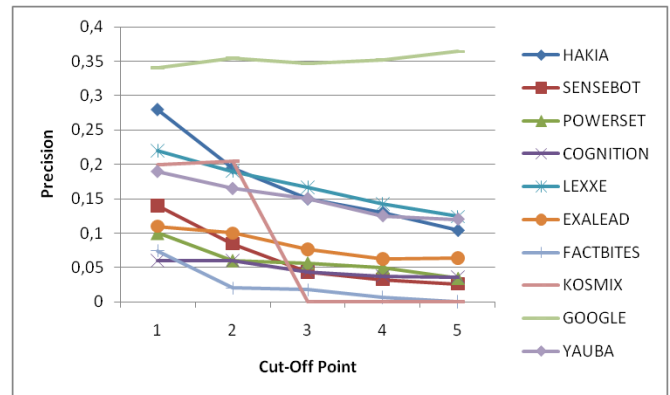

Fig. 1. Mean precision ratios of semantic search engines and Google

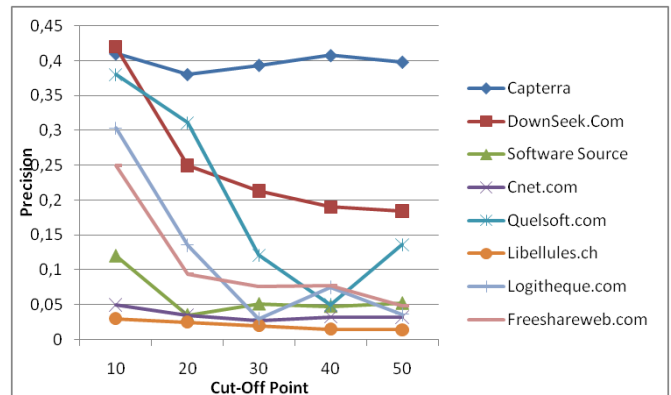

Fig. 2. Mean precision ratios of software component search engines

TABLE XI
NUMBER OF RELEVANT DOCUMENTS RETRIEVED BY SOFTWARE COMPONENT SEARCH ENGINES

| Query Number | Capterra | Downseek | Software Source | Cnet | Quelsoft | Libellules | Logitheque | Freeshareweb |
|---|---|---|---|---|---|---|---|---|
| Q1 | 24 | 24 | 1 | 0 | 9 | 2 | 18 | 24 |
| Q2 | 5 | 9 | 0 | 0 | 0 | 0 | 2 | 0 |
| Q3 | 0 | 3 | 2 | 0 | 0 | 0 | 0 | 0 |
| Q4 | 3 | 6 | 0 | 3 | 2 | 3 | 1 | 1 |
| Q5 | 21 | 14 | 1 | 2 | 6 | 0 | 8 | 1 |
| Q6 | 5 | 5 | 1 | 2 | 1 | 0 | 5 | 0 |
| Q7 | 4 | 1 | 0 | 1 | 4 | 0 | 1 | 9 |
| Q8 | 47 | 5 | 8 | 1 | 17 | 2 | 4 | 1 |
| Q9 | 50 | 5 | 3 | 3 | 2 | 0 | 1 | 0 |
| Q10 | 40 | 1 | 12 | 4 | 10 | 0 | 0 | 0 |
| Total Relevant | 199 | 93 | 28 | 16 | 61 | 7 | 40 | 36 |
| Total Retrieved | 500 | 451 | 325 | 500 | 209 | 300 | 119 | 98 |
| Average | 19,9 | 9,3 | 2,8 | 1,6 | 6,1 | 0,7 | 4 | 3,6 |

The second class includes the semantic search engines with the lowest ratios, which are Sensebot, Exalead, Powerset, Cognition, Factbites and Kosmix. Two factors can justify this classification: 1) Nature of searched document which describes the nature of Web documents (structured or unstructured) indexed and retrieved by the search engine, 2) Portals it relies on which describes the repositories explored by the semantic search engine to find information. In fact, in the state of practice, descriptions of COTS components are provided as heterogeneous and unstructured web pages. Hakia, Lexxe, and Exalead have the highest precision ratios since they search for results in the entire Web. However, Powerset and Cognition search only for structured Web documents. That's why, these latter don't succeed in indexing and retrieving COTS components. On the other hand, Sensebot is based on the summarization of search results. Sensebot can be useful for preparing summaries about a general topic; users can benefit from it when they are trying to understand a concept or a particular area of knowledge. But Sensebot is useless when users look for particular information; it presents really poor results. As a consequence, Powerset, Cognition, and Sensebot are absolutely inappropriate for searching COTS components. Regarding Kosmix, Exalead, and Factbites, even though these search engines index structured and unstructured documents, they also have the lowest precision ratios. As a matter of fact, Kosmix is based on metasearching of Google and Bing; it just presents some search results taken from these two search engines.

On the other hand, we can notice that although Google is a keyword-based search engine and does not use semantics in his search process, it has higher precision ratios than all semantic search engines. However, its precision ratios are low. In fact, the average precisions of Google are between 34% and 36%. In other words, Google returns about 7 non relevant in 10 retrieved documents. Consequently, even Google is inappropriate for searching COTS components marketed on the Web.

Regarding software components search engines, Downseek has the highest precision ratio at cut-off point 10 (42%). Whereas, Capterra has the highest precision ratios at cut-off points 20, 30, 40, and 50 with respectively 38%, 39%, 40%, and 39%. In addition, we notice that only precision ratios of Downseek and Libellules decreased for all cut-off points.

If we examine both precision ratios of semantic search engines, software components search engines, and Google, Capterra has higher precision ratios than Google and all semantic search engines. Besides, Google has higher precision ratios on all cut-off points than 5 software components search engines, namely Software Source, Cnet, Libellules, Logitheque, and Freeshareweb. In the same way, Google has higher precision ratios than Downseek and Quelsoft at cut-off points 20, 30, 40, and 50.

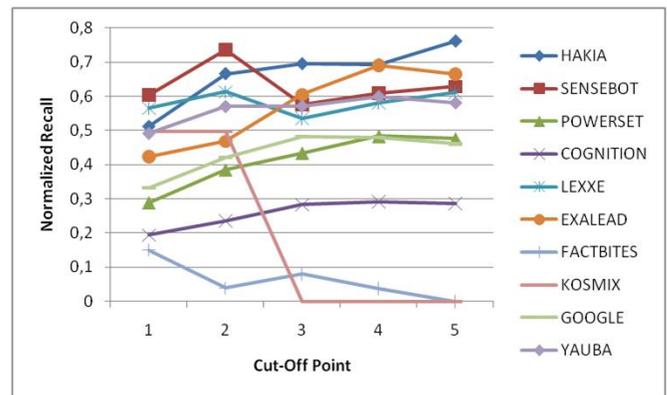

Fig. 4. Mean normalized recall ratios of software component search engines

As a conclusion, we first notice that the assessed search engines have low precision ratios that do not encourage COTS component consumers to use them to identify and retrieve COTS components. Secondly, although the semantic search engines are meaning-based and the component-specific search engines are specialized ones, they have lower precision ratios than Google.

*Normalized Recall Ratios*

The normalized recall ratio measures if search engines display relevant documents in the top ranks of the retrieval outputs. If a search engine could not retrieve any documents for a search query, the normalized recall value for that query will be zero [7]. Mean normalized recall values of the assessed search engines in various cut-off points (for first 5, 10, 15, and 20 documents retrieved) are shown in Fig. 3 and Fig. 4.

As illustrated in Fig. 3, Hakia has the highest (76%) normalized recall value at cut-off point 50. Hakia's normalized recall ratio increased gradually at all cut-off points. The normalized recall ratios of Factbites and Cognition are the lowest as it could not retrieve any relevant document for 6 of 10 queries. Kosmix has 0% as normalized recall value at cut-off points 30, 40, and 50 since total of retrieved documents per search query does not exceed 18 documents.

Moreover, 5 semantic search engines, as illustrated by Fig. 3, have higher performances than Google for displaying relevant documents retrieved in the top ranks of the retrieval output, namely Exalead, Yauba, Lexxe, Hakia, and Sensebot. On the other hand, there was no statistically meaningful difference between normalized recall values of search engines on all cut-off points. In other words, none of the search engines displayed relevant documents in distinguishably higher ranks than others in general.

Regarding software component search engines, Downseek has the highest performance for displaying relevant documents retrieved in the top ranks of the retrieval output as shown in Fig. 4. Quelsoft's normalized recall ratio was approximately the same as that of Downseek at cut-off points 10 and 20. Similarly, normalized recall ratios of Cnet and Logitheque on the one hand, and Software source and Freeshareweb, on the other hand, were approximately the same at cut-off 20.

If we examine both normalized recall ratios of semantic search engines, software components search engines, and Google, we notice that Sensebot has higher normalized recall ratios on all cut-off points than 7 software component search engines, namely Capterra, Software Source, Cnet, Libellules, Logitheque, Quelsoft, and Freeshareweb. Google has higher normalized recall ratios on all cut-off points than 6 software component search engines. Besides, Google has higher normalized recall ratio than Quelsoft at cut-off points 30, 40, and 50. Only Downseek, as a software component search engine, has higher performance than Google for displaying relevant documents retrieved in the top ranks.

*The Relation Between Precision and Normalized Recall*

The relationship between precision and normalized recall ratios was not statistically significant. In other words, the normalized recall ratios of the semantic and software component search engines were not high when their precision ratios were high. Our empirical evaluation revealed that the number of relevant documents in the retrieval output does not tend to decrease as one goes down in the result list. For example, Downseek has 42% as precision ratio at cut-off point 10 and 52% as normalized recall ratio. However, the same search engine has 59% as normalized recall ratio at cut-off value 50 when its precision ratio was 18%. Likewise, Libellules precision ratios decrease from 3% at cut-off point 10 to 1% at cut-off point 50. Nevertheless, its normalized recall ratios increase from 9% at cut-off point 10 to 21% at cut-off point 50.

We observed the same finding in the semantic search engines. In fact, Hakia precision ratios decreased from 28% to 10% while its normalized recall ratios increased from 51% to 76% at cut-off points 10 and 50. Similarly, Yauba precision ratios decreased from 19% to 12% whereas its normalized recall ratios increased from 49% to 58% at cut-off points 10 and 50.

On the other hand, Google is the only assessed search engine whose normalized recall ratios were high when the precision ratios were high. For instance, when Google precision ratios were 34% and 35% at cut-off points 10 and 40, its normalized recall ratios were 33% and 47%, respectively.

## I. CONCLUSION AND WORK IN PROGRESS

In this paper, an investigative evaluation on search performance of keyword-based, specialized and semantic search engines is detailed in the context of COTS component identification.

Regarding components specialized search engines, our study revealed that although they are a good way of getting started, since they manage small software components indexes which make easier the searching task. However, many shortcomings have been detected such as the absence of refinement options, the non integration of the semantic aspect in order to improve the results relevance and the lack of personalization.

Similarly, we detected several limitations following the theoretical study of the semantic search engines and Google. In fact, the majority of these search engines do not allow users to refine their queries in order to take into account the specific characteristics of COTS components (target platform, vendor, etc.). Besides, none of the assessed software component search engines focus on representation and exploitation of users' preferences and intentions during search process.

On the other hand, our empirical evaluation confirms that the results of the assessed search engines are not encouraging and very far from satisfying the expectations of CBD users. In fact, it was found that Google, Hakia, Yauba, Lexxe, Exalead, and Downseek retrieved at least one relevant document for all queries. However, the remainder of the assessed search engines could not retrieve any relevant document for at least one query.

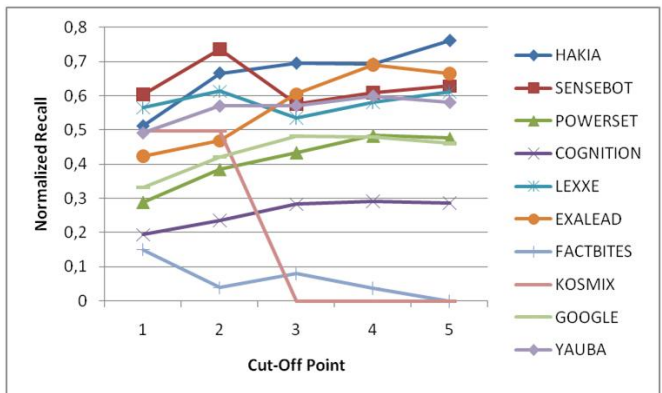

Fig. 3. Mean normalized recall ratios of semantic search engines and Google

In terms of overall performance, Google retrieved more relevant documents compared to the rest of the assessed search engines. Furthermore, it has higher precision ratios than all semantic search engines and 5 software component search engines. However, Google precision ratios were low. In fact, its average precisions are between 34% and 36%.

Generally, precision ratios of search engines decreased with increased cut-off point values. However, it was seen from the results that the performance of search engines to display relevant documents in the top ranks, is no better than their relevant document retrieval.

Finally, it was seen that the semantic search was low regardless of the type of the search engine used. Therefore, search engines need to improve their systems, taking into consideration the importance of the role semantic search can play in helping users getting precise information from the Web with minimal effort.

As work in progress, we are now focusing on the design of an intelligent user-centered search engine for COTS components marketed on the Web. The proposed search engine integrates semantic search technologies and knowledge about COTS components as well as users into a single framework in order to provide the most appropriate COTS components for user's needs. To integrate the intelligent and user-centered aspects in our COTS component search engine, we use two kinds of ontologies. An evolvable COTS component ontology describes concepts and relations between these concepts appearing in a COTS component and unifies the heterogeneous COTS descriptions available on the Web. On the other hand, application domains ontologies represent and store knowledge about COTS specific domains such as security, ERP (Enterprise Resource Planning), CRM (Customer Relationship Management), e-Commerce, etc. We also anticipate the integration of a user model that represents preferences and interest domains of the user. The proposed user model is used in each step of the search process. For instance, the user model is used to set weights to the keywords in the query in order to indicate the relative interest of the user for each of these keywords. It is also used to limit the search space for only COTS components corresponding to the user domain of interest.